\documentclass[aps,prl,twocolumn,showpacs,superscriptaddress,groupedaddress]{revtex4}  
\usepackage{graphicx}  
\usepackage{dcolumn}   
\usepackage{bm}        
\usepackage{amssymb}   

\def\MUMU{$\mu^+\mu^-$}
\def\BS{$B_s^0$}
\def\BU{$B^+$}
\def\BD{$B^0$}
\def\BSJPSIPHI{$B_s^0 \rightarrow J/\psi\phi$}
\def\BRBSMUMU{${\cal B}(B_s^0 \rightarrow \mu^+ \mu^-)$}
\def\BSKK{$B_s^0 \rightarrow K^+ K^-$}
\def\BUJPSIK{$B^+ \rightarrow J/\psi (\mu^+ \mu^-) K^+$}

\def\BDJPSIK{$B^0 \rightarrow J/\psi (\mu^+ \mu^-) K^{*0} (K^+\pi^-)$}

\def\BHH{$B \rightarrow h^+ h'^-$}
\def\BSMUMU{$B_s^0 \rightarrow \mu^+ \mu^-$}
\def\JPSIMUMU{$J/\psi \rightarrow \mu^+ \mu^-$}
\def\BMUD{$B \rightarrow \mu\nu D^0$}
\def\DKPI{$D^0 \rightarrow K^+ \pi^-$}

\def\INVFB{${\rm fb}^{-1}$}
\def\PPBAR{$p{\bar p}$}
\def\SQRTS{$\sqrt{s} = 1.96~{\rm TeV}$}
\def\RUNII{Run~${\rm I\hspace{-0.1em}I}$}
\def\RUNIIA{Run~${\rm I\hspace{-0.1em}Ia}$}
\def\RUNIIB{Run~${\rm I\hspace{-0.1em}Ib}$}

\def\FULLEXPLIMITBR{${\cal B}(B_s^0 \rightarrow \mu^+ \mu^-) < 4.0 \times 10^{-8}  (3.2 \times 10^{-8}) $}

\def\OBSLIMITBR{${\cal B}(B_s^0 \rightarrow \mu^+ \mu^-) < 5.1 \times 10^{-8}$}

\def\FULLOBSLIMITBR{${\cal B}(B_s^0 \rightarrow \mu^+ \mu^-) < 5.1 \times 10^{-8}  (4.2 \times 10^{-8}) $}

\def\FULLEXPLIMITBRA{${\cal B}(B_s^0 \rightarrow \mu^+ \mu^-) < 8.5 \times 10^{-8} (6.8 \times 10^{-8})$}

\def\FULLOBSLIMITBRA{${\cal B}(B_s^0 \rightarrow \mu^+ \mu^-) < 8.2 \times 10^{-8} (6.5 \times 10^{-8})$}

\def\FULLEXPLIMITB{$4.6 \times 10^{-8} (3.7 \times 10^{-8})$}

\def\FULLOBSLIMITB{$6.5 \times 10^{-8} (5.3 \times 10^{-8})$}

\def\BNN{$\beta$}

\begin{document}


\hspace{5.2in} \mbox{FERMILAB-PUB-10-202-E}

\title{Search for the rare decay {\boldmath \BSMUMU}}

\affiliation{Universidad de Buenos Aires, Buenos Aires, Argentina}
\affiliation{LAFEX, Centro Brasileiro de Pesquisas F{\'\i}sicas, Rio de Janeiro, Brazil}
\affiliation{Universidade do Estado do Rio de Janeiro, Rio de Janeiro, Brazil}
\affiliation{Universidade Federal do ABC, Santo Andr\'e, Brazil}
\affiliation{Instituto de F\'{\i}sica Te\'orica, Universidade Estadual Paulista, S\~ao Paulo, Brazil}
\affiliation{Simon Fraser University, Vancouver, British Columbia, and York University, Toronto, Ontario, Canada}
\affiliation{University of Science and Technology of China, Hefei, People's Republic of China}
\affiliation{Universidad de los Andes, Bogot\'{a}, Colombia}
\affiliation{Charles University, Faculty of Mathematics and Physics, Center for Particle Physics, Prague, Czech Republic}
\affiliation{Czech Technical University in Prague, Prague, Czech Republic}
\affiliation{Center for Particle Physics, Institute of Physics, Academy of Sciences of the Czech Republic, Prague, Czech Republic}
\affiliation{Universidad San Francisco de Quito, Quito, Ecuador}
\affiliation{LPC, Universit\'e Blaise Pascal, CNRS/IN2P3, Clermont, France}
\affiliation{LPSC, Universit\'e Joseph Fourier Grenoble 1, CNRS/IN2P3, Institut National Polytechnique de Grenoble, Grenoble, France}
\affiliation{CPPM, Aix-Marseille Universit\'e, CNRS/IN2P3, Marseille, France}
\affiliation{LAL, Universit\'e Paris-Sud, CNRS/IN2P3, Orsay, France}
\affiliation{LPNHE, Universit\'es Paris VI and VII, CNRS/IN2P3, Paris, France}
\affiliation{CEA, Irfu, SPP, Saclay, France}
\affiliation{IPHC, Universit\'e de Strasbourg, CNRS/IN2P3, Strasbourg, France}
\affiliation{IPNL, Universit\'e Lyon 1, CNRS/IN2P3, Villeurbanne, France and Universit\'e de Lyon, Lyon, France}
\affiliation{III. Physikalisches Institut A, RWTH Aachen University, Aachen, Germany}
\affiliation{Physikalisches Institut, Universit{\"a}t Freiburg, Freiburg, Germany}
\affiliation{II. Physikalisches Institut, Georg-August-Universit{\"a}t G\"ottingen, G\"ottingen, Germany}
\affiliation{Institut f{\"u}r Physik, Universit{\"a}t Mainz, Mainz, Germany}
\affiliation{Ludwig-Maximilians-Universit{\"a}t M{\"u}nchen, M{\"u}nchen, Germany}
\affiliation{Fachbereich Physik, Bergische  Universit{\"a}t Wuppertal, Wuppertal, Germany}
\affiliation{Panjab University, Chandigarh, India}
\affiliation{Delhi University, Delhi, India}
\affiliation{Tata Institute of Fundamental Research, Mumbai, India}
\affiliation{University College Dublin, Dublin, Ireland}
\affiliation{Korea Detector Laboratory, Korea University, Seoul, Korea}
\affiliation{CINVESTAV, Mexico City, Mexico}
\affiliation{FOM-Institute NIKHEF and University of Amsterdam/NIKHEF, Amsterdam, The Netherlands}
\affiliation{Radboud University Nijmegen/NIKHEF, Nijmegen, The Netherlands}
\affiliation{Joint Institute for Nuclear Research, Dubna, Russia}
\affiliation{Institute for Theoretical and Experimental Physics, Moscow, Russia}
\affiliation{Moscow State University, Moscow, Russia}
\affiliation{Institute for High Energy Physics, Protvino, Russia}
\affiliation{Petersburg Nuclear Physics Institute, St. Petersburg, Russia}
\affiliation{Stockholm University, Stockholm and Uppsala University, Uppsala, Sweden }
\affiliation{Lancaster University, Lancaster LA1 4YB, United Kingdom}
\affiliation{Imperial College London, London SW7 2AZ, United Kingdom}
\affiliation{The University of Manchester, Manchester M13 9PL, United Kingdom}
\affiliation{University of Arizona, Tucson, Arizona 85721, USA}
\affiliation{University of California Riverside, Riverside, California 92521, USA}
\affiliation{Florida State University, Tallahassee, Florida 32306, USA}
\affiliation{Fermi National Accelerator Laboratory, Batavia, Illinois 60510, USA}
\affiliation{University of Illinois at Chicago, Chicago, Illinois 60607, USA}
\affiliation{Northern Illinois University, DeKalb, Illinois 60115, USA}
\affiliation{Northwestern University, Evanston, Illinois 60208, USA}
\affiliation{Indiana University, Bloomington, Indiana 47405, USA}
\affiliation{Purdue University Calumet, Hammond, Indiana 46323, USA}
\affiliation{University of Notre Dame, Notre Dame, Indiana 46556, USA}
\affiliation{Iowa State University, Ames, Iowa 50011, USA}
\affiliation{University of Kansas, Lawrence, Kansas 66045, USA}
\affiliation{Kansas State University, Manhattan, Kansas 66506, USA}
\affiliation{Louisiana Tech University, Ruston, Louisiana 71272, USA}
\affiliation{University of Maryland, College Park, Maryland 20742, USA}
\affiliation{Boston University, Boston, Massachusetts 02215, USA}
\affiliation{Northeastern University, Boston, Massachusetts 02115, USA}
\affiliation{University of Michigan, Ann Arbor, Michigan 48109, USA}
\affiliation{Michigan State University, East Lansing, Michigan 48824, USA}
\affiliation{University of Mississippi, University, Mississippi 38677, USA}
\affiliation{University of Nebraska, Lincoln, Nebraska 68588, USA}
\affiliation{Rutgers University, Piscataway, New Jersey 08855, USA}
\affiliation{Princeton University, Princeton, New Jersey 08544, USA}
\affiliation{State University of New York, Buffalo, New York 14260, USA}
\affiliation{Columbia University, New York, New York 10027, USA}
\affiliation{University of Rochester, Rochester, New York 14627, USA}
\affiliation{State University of New York, Stony Brook, New York 11794, USA}
\affiliation{Brookhaven National Laboratory, Upton, New York 11973, USA}
\affiliation{Langston University, Langston, Oklahoma 73050, USA}
\affiliation{University of Oklahoma, Norman, Oklahoma 73019, USA}
\affiliation{Oklahoma State University, Stillwater, Oklahoma 74078, USA}
\affiliation{Brown University, Providence, Rhode Island 02912, USA}
\affiliation{University of Texas, Arlington, Texas 76019, USA}
\affiliation{Southern Methodist University, Dallas, Texas 75275, USA}
\affiliation{Rice University, Houston, Texas 77005, USA}
\affiliation{University of Virginia, Charlottesville, Virginia 22901, USA}
\affiliation{University of Washington, Seattle, Washington 98195, USA}
\author{V.M.~Abazov} \affiliation{Joint Institute for Nuclear Research, Dubna, Russia}
\author{B.~Abbott} \affiliation{University of Oklahoma, Norman, Oklahoma 73019, USA}
\author{M.~Abolins} \affiliation{Michigan State University, East Lansing, Michigan 48824, USA}
\author{B.S.~Acharya} \affiliation{Tata Institute of Fundamental Research, Mumbai, India}
\author{M.~Adams} \affiliation{University of Illinois at Chicago, Chicago, Illinois 60607, USA}
\author{T.~Adams} \affiliation{Florida State University, Tallahassee, Florida 32306, USA}
\author{G.D.~Alexeev} \affiliation{Joint Institute for Nuclear Research, Dubna, Russia}
\author{G.~Alkhazov} \affiliation{Petersburg Nuclear Physics Institute, St. Petersburg, Russia}
\author{A.~Alton$^{a}$} \affiliation{University of Michigan, Ann Arbor, Michigan 48109, USA}
\author{G.~Alverson} \affiliation{Northeastern University, Boston, Massachusetts 02115, USA}
% ...
%
%
\collaboration{The D0 Collaboration\footnote{with visitors from
$^{a}$Augustana College, Sioux Falls, SD, USA,
$^{b}$The University of Liverpool, Liverpool, UK,
$^{c}$SLAC, Menlo Park, CA, USA,
$^{d}$ICREA/IFAE, Barcelona, Spain,
$^{e}$Centro de Investigacion en Computacion - IPN, Mexico City, Mexico,
$^{f}$ECFM, Universidad Autonoma de Sinaloa, Culiac\'an, Mexico,
and 
$^{g}$Universit{\"a}t Bern, Bern, Switzerland.%
}} \noaffiliation
\vskip 0.25cm
\date{June 16, 2010}

\begin{abstract}

  We present the results of a search for the flavor changing neutral current decay \BSMUMU\ using 6.1~\INVFB\ of \PPBAR\ collisions at \SQRTS\ collected by the D0 experiment at the Fermilab Tevatron Collider.
  The observed number of \BS\ candidates is consistent with background expectations.
  The resulting upper limit on the branching fraction is \OBSLIMITBR\ at the 95\%~C.L.
  This limit is a factor of 2.4 better than that of the previous D0 analysis and the best limit to date.

\end{abstract}



\pacs{12.15.Mn, 13.20.He, 14.40.Nd}
\maketitle


The standard model (SM) provides an accurate description of current observations in
high energy physics experiments, in particular precision electroweak 
measurements and flavor physics observables.
A flavor changing neutral current (FCNC) process is an apparent transition between quarks of different flavor but equal charge.
In the SM, the FCNC processes are forbidden at first order.
They can occur at second order only through Glashow-Iliopoulos-Maiani (GIM)~\cite{Glashow:1970gm}
suppressed box and penguin diagrams.
\begin{figure}
  \begin{center}
  \includegraphics[scale=0.4]{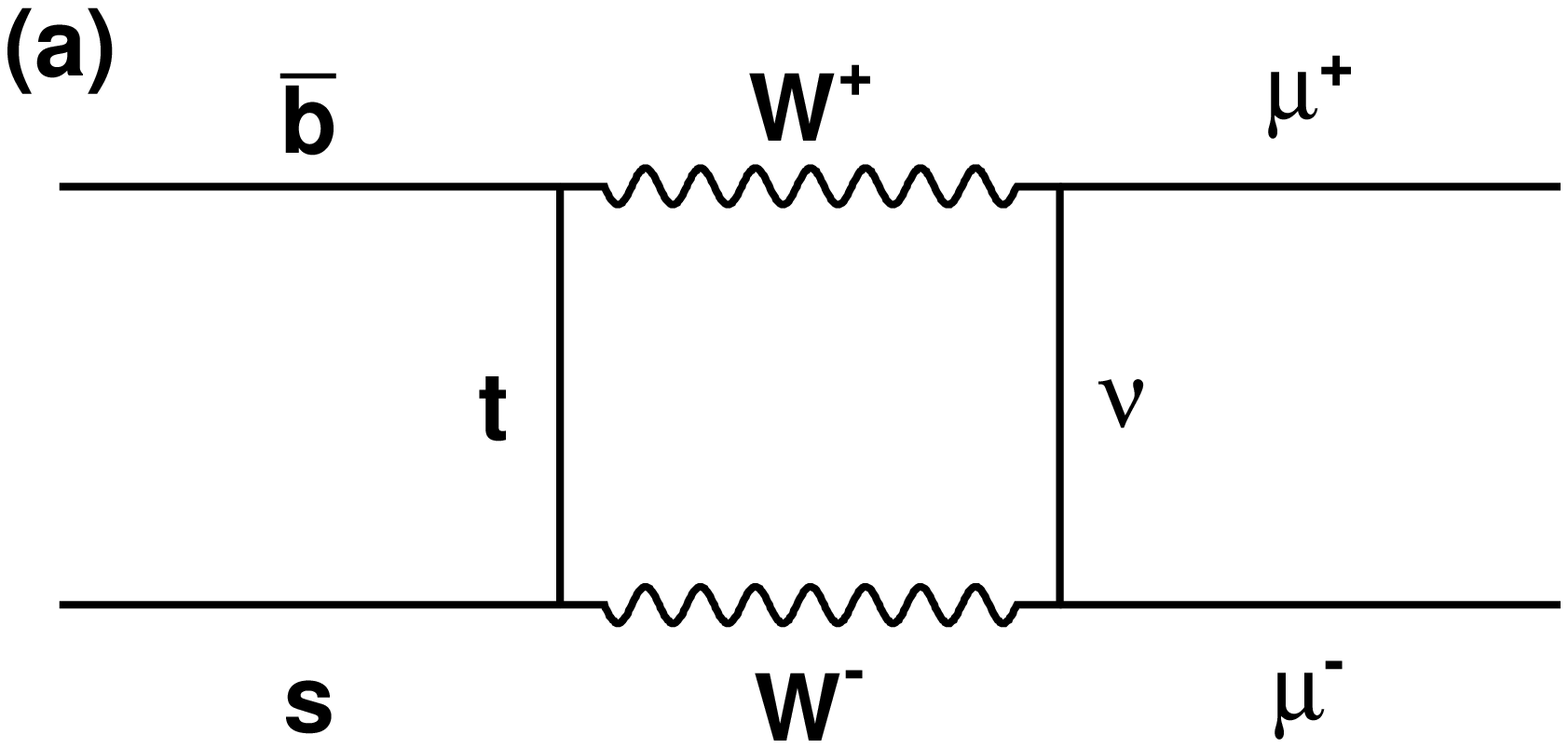}
  \includegraphics[scale=0.4]{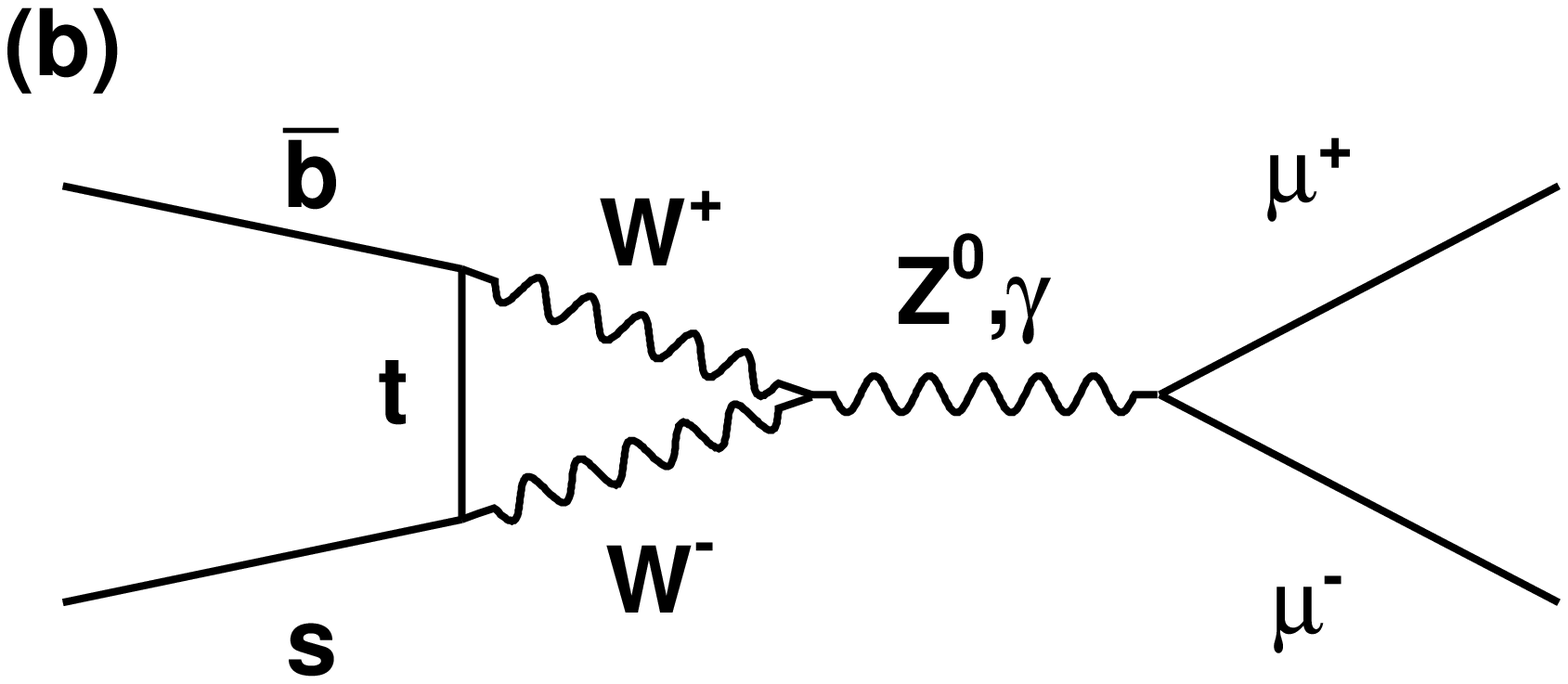}
  \caption{\label{fig:feynman_diagrams} Examples of Feynman diagrams for FCNC decays: (a) box diagram, (b) penguin diagram.}
  \end{center}
\end{figure}
The decay \BSMUMU~\cite{charge_conjugation} is an example of such a process, as shown in Fig.~\ref{fig:feynman_diagrams}.
Unlike other FCNC decays this decay rate is further suppressed by helicity factors in the \MUMU\ final state~\cite{Buchalla:1993bv}.
The SM expectation for the branching fraction of this decay is  
$(3.6 \pm 0.3) \times 10^{-9}$~\cite{Buras:2009us}.
The decay amplitude for \BSMUMU\ can be enhanced by 
several orders of magnitude in some extensions of the SM.
For example,  in some supersymmetric models such as the minimal supersymmetric standard model this decay can occur through the mediation of superpartners of the SM intermediate particles as well as particles from the extended Higgs sector.
This extended contribution becomes larger if the value of $\tan\beta$, the ratio of the vacuum expectation values of the two neutral Higgs fields,
is large~\cite{Guadagnoli:2009ze,Allanach:2009ne,Parry:2006mv,Lunghi:2006uf,Parry:2005fp,Choudhury:1998ze}.
Similarly, in some supersymmetric models with $R$-parity violating couplings~\cite{Arnowitt:2002cq},
this enhancement can be present even in the regime of low $\tan\beta$. 
Improved limits on the branching fraction of this decay can be used to set limits on the parameter space  of supersymmetric models and other new theories.
Since the predicted rate for this process in the SM is beyond the current experimental sensitivity at the Tevatron,
the observation of this decay would necessarily imply physics beyond the SM.
Similar annihilation topologies have also been studied for $B^0$ and $B^+$ decays~\cite{Aubert:2007xj,:2008gx,Ikado:2006un,:2008ch,cdf:2007kv}.

In this Letter, we report on a search for the rare decay
\BSMUMU\ using $6.1~{\rm fb}^{-1}$ of integrated luminosity collected by the D0 detector.
Presently, the best experimental bound for the branching fraction of
${\cal B}(B_s^0 \to \mu^+ \mu^-)<5.8\times 10^{-8}$ at the 95\%~C.L. is given by the CDF Collaboration~\cite{cdf:2007kv}.
Our previous result for this search was based on  $1.3~{\rm fb}^{-1}$ of integrated luminosity and set a bound for the branching fraction ${\cal B}(B_s^0 \to \mu^+ \mu^-)<1.2\times 10^{-7}$ at the 95\%~C.L.~\cite{Abazov:2007iy}.

The D0 detector~\cite{run2det} has a central tracking system, consisting of a 
silicon microstrip tracker (SMT)~\cite{smt} and a central fiber tracker (CFT), 
both located within a 2~T superconducting solenoidal 
magnet, with designs optimized for tracking and 
vertexing at pseudorapidities $|\eta| \lesssim  3$ and $|\eta|  \lesssim  2.5$, respectively, where
$\eta = - \ln \left[ \tan{ (\theta / 2) } \right]$, and  $\theta$ is the polar angle with respect to the proton beam direction.
An outer muon system, covering $|\eta| \lesssim 2$, 
consists of a layer of tracking detectors and scintillation trigger 
counters in front of 1.8~T toroids, followed by two similar layers 
after the toroids~\cite{run2muon}. 
The trigger and data acquisition systems are designed to accommodate the high instantaneous luminosity of the Tevatron \RUNII\ that started in 2001.
In summer 2006, the SMT detector was upgraded by inserting an additional layer of silicon microstrip detectors, Layer~0~\cite{l0}, close to the beampipe.
The data-taking period before the Layer~0 installation is referred to as \RUNIIA, and the period afterwords is referred to as \RUNIIB.
The two data sets are analyzed separately.


All data collected up to June 2009 are included in this analysis.
The integrated luminosities for the \RUNIIA\ and \RUNIIB\ data sets are $1.3~{\rm fb}^{-1}$ and $4.8~{\rm fb}^{-1}$, respectively.
Events are recorded using a set of single muon triggers, dimuon triggers,
and triggers that select \PPBAR\ interactions based on energy depositions in the calorimeter.
\BSMUMU candidates are formed from pairs of oppositely charged muons identified by
extrapolating tracks reconstructed in the central tracking detectors to the muon detectors, and matching them with
information from the muon system.
The muon selection has been updated with respect to the previous analysis~\cite{Abazov:2007iy}, yielding 10\% higher acceptance while keeping the fraction of misidentified muons below 0.5\%.
Each muon is required to have a transverse momentum $p_T^{\mu}  \geq  1.5~{\rm GeV}$, and to have hits in at least two layers of both the CFT and the SMT.
The \BS\ candidate is required to have a reconstructed three-dimensional (3D) decay vertex displaced from the interaction point with
a transverse decay length significance $L_{T}/\sigma_{L_{T}} \geq 3$ to reduce prompt dimuon background, where $L_{T}=\vec{l}_{T} \cdot \vec{p}_T^{B} / |\vec{p}_T^{B}|$.
The vectors $\vec{l}_{T}$ and $\vec{p}_T^{B}$ are, respectively, the vector from the interaction point to the decay point and the transverse momentum vector of the \BS\ meson in the transverse plane.
The \PPBAR\ interaction vertex is found for each event using a beam-spot constrained fit as described in~\cite{Abdallah:2002xm}.
Events are selected if the reconstructed invariant dimuon mass, $m_{\mu\mu}$, is between $4.0~{\rm GeV}$ and $7.0~{\rm GeV}$.

\begin{figure}
  \begin{center}
  \includegraphics[scale=0.43]{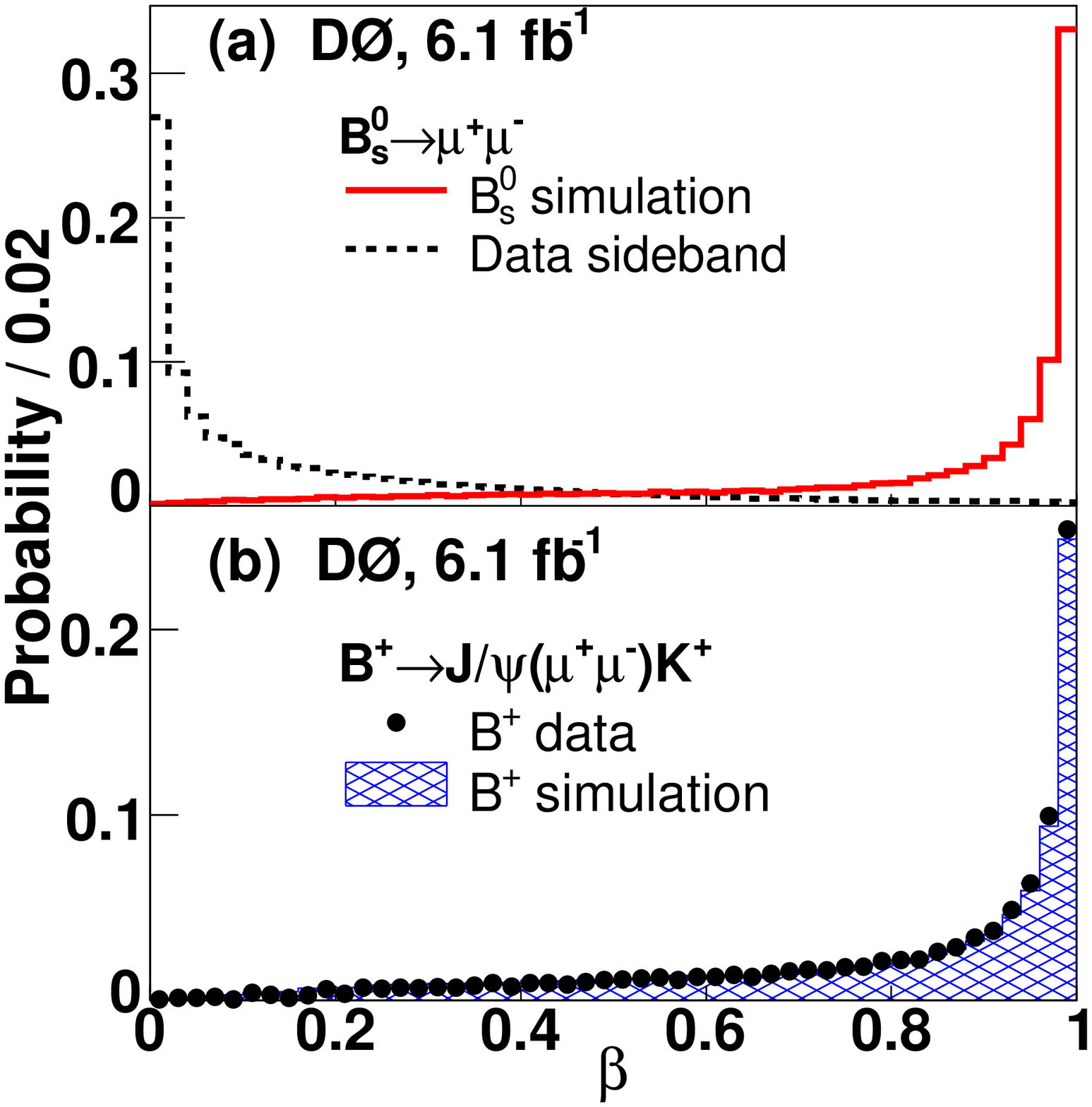}
  \caption{\label{fig:nnout} Distributions of \BNN: (a) \BSMUMU\ signal and sideband events, (b) \BUJPSIK\ data and simulation.}
  \end{center}
\end{figure}

To further suppress the background we use the following discriminating variables:
the transverse momentum of the \BS\ candidate $p_T^B$, the pointing angle, $L_{T}/\sigma_{L_{T}}$, the decay vertex fit $\chi^2$, the smaller impact parameter significance ($\delta/\sigma_\delta$)  of the two muons, ${\rm min}(\delta/\sigma_\delta)$, and the smaller $p_T^{\mu}$ of the two muons, ${\rm min}(p_T^{\mu})$.
The pointing angle is defined to be the 3D opening angle between the $B^0_s$ meson momentum vector and the displacement vector from the interaction to the dimuon vertex.
The impact parameter $\delta$ is defined to be the distance of closest approach of the track to the interaction point in the transverse plane, and $\sigma_\delta$ is its uncertainty.
We use a Bayesian Neural Network (BNN)~\cite{bnn1,bnn3} multivariate classifier with the above variables to distinguish signal events from background.
The BNN is trained using background events sampled from the sideband regions ($4.5~{\rm GeV} \leq m_{\mu\mu} \leq 5.0~{\rm GeV}$ and $5.8~{\rm GeV} \leq m_{\mu\mu} \leq 6.5~{\rm GeV}$)
and simulated signal events.
To simulate the \BS\ signal, we generate  Monte Carlo events using the {\sc pythia}~\cite{pythia} event generator, interfaced with the {\sc evtgen}~\cite{evtgen} decay package.We simulate the detector response using {\sc geant}~\cite{geant}.
Multiple interactions are modeled by overlaying randomly triggered data events on top of the simulated hits in the detector.
The distributions of the BNN output \BNN\ for the \BS\ signal and the sideband events as well as the \BUJPSIK\ control sample are shown in Fig.~\ref{fig:nnout}.
We define the \BSMUMU\ signal region to be $0.9 \leq {\rm \beta} \leq 1.0$ and $5.0~{\rm GeV} \leq m_{\mu\mu} \leq 5.8~{\rm GeV}$ where there is a clear separation between signal and background.
This region is determined by optimizing the expected sensitivity of the search. 
We prepare two-dimensional (2D) histograms of $m_{\mu\mu}$~vs.~$\beta$ dividing the signal region into several bins to improve the sensitivity relative to using a single bin.

The dominant source of background dimuon events is from decays of heavy flavor hadrons in $b\bar{b}$ or $c\bar{c}$ production.
To study this background contribution, we generate inclusive dimuon Monte Carlo samples with {\sc pythia} generic QCD processes
that include all $b{\bar b}$ or $c{\bar c}$ production processes.
The dimuon background events can be categorized by two types:
(i) $B(D) \to \mu^+ \nu X, {\bar B}(\bar{D}) \to \mu^- \bar{\nu} X'$ double semileptonic decays where the two muons originate from different $b(c)$ quarks,
yielding dimuon masses distributed over the entire signal region, 
and (ii) $B \to \mu^+ \nu \bar{D}, \bar{D} \to \mu^- \bar{\nu} X$ sequential semileptonic decays,
resulting in $m_{\mu\mu}$ predominantly below the $B$ hadron mass.
The simulated dimuon mass distributions for both background sources after requiring $\beta \geq 0.8$ are parametrized using an exponential function
to estimate the number of background events in the signal region
after fitting the dimuon mass in the data sideband regions, $4.0~{\rm GeV} \leq m_{\mu\mu} \leq 5.0~{\rm GeV}$ and $6.0~{\rm GeV} \leq m_{\mu\mu} \leq 7.0~{\rm GeV}$,
in each \BNN\ bin.
The uncertainty on this background estimate is dominated by the statistical uncertainty of the sideband sample ($10$--$35\%$).
In addition, we consider background contributions from \BD\ and \BS\ decays \BHH, where $h^+$ and $h'^-$ represent a charged kaon or pion.
The muon identification efficiency and the fractions of pions and kaons misidentified as muons are evaluated using samples of \JPSIMUMU\ and \DKPI\ in \BMUD\ decays.
\BSKK\ decay is the largest contribution in the \BHH\ backgrounds and that is expected to be $0.13 \pm 0.10$ events for \RUNIIA\ and $0.36 \pm 0.27$ events for \RUNIIB\ in the signal region, where the uncertainty is dominated by the statistical uncertainty on the fraction of misidentification.
The \BHH\ background contribution is thus found to be negligible (see below).

The branching fraction \BRBSMUMU\ is computed by normalizing the number of events, $N(B_s^0)$, to the number of reconstructed \BUJPSIK\ events, $N(B^+)$:
\begin{equation}
   {\cal B} \left( B_s^0 \rightarrow \mu^+ \mu^- \right) =  \frac{ N \left( B_s^0 \right)}{N(B^+ )} \cdot \frac{\epsilon_{B^+}}{\epsilon_{B_s^0}} \cdot \frac{f_u}{f_s} \cdot {\cal B} \left( B^+ \right),
   \label{eq:brcalc}
\end{equation}
where the parameters $\epsilon_{B^+}$ and $\epsilon_{B_s^0}$ are the reconstruction efficiencies for \BUJPSIK\ and \BSMUMU, respectively.
They are estimated from simulations.
We use ${\cal B}(B^+) = {\cal B}\left(B^+ \rightarrow J/\psi K^+\right) \times {\cal B}\left( J/\psi \rightarrow \mu^+\mu^-\right) = (5.97 \pm 0.22) \times 10^{-5}$~\cite{pdg2008} and the ratio of $B$-hadron production fractions  $f_u/f_s = 3.86 \pm 0.59$~\cite{pdg2006}.
The simulated mass resolution of the D0 detector for the \BSMUMU\ is $\approx120~{\rm MeV}$
and is therefore insufficient to readily separate \BS\ from \BD\ leptonic decays.
In this analysis, we assume that there are no contributions from $B^0 \rightarrow \mu^+ \mu^-$ decays,
since this decay is suppressed by $|V_{td}/V_{ts}|^2 \approx 0.04$~\cite{Cabibbo,Kobayashi:1973fv}.

\begin{figure}
  \begin{center}
   \includegraphics[scale=0.43]{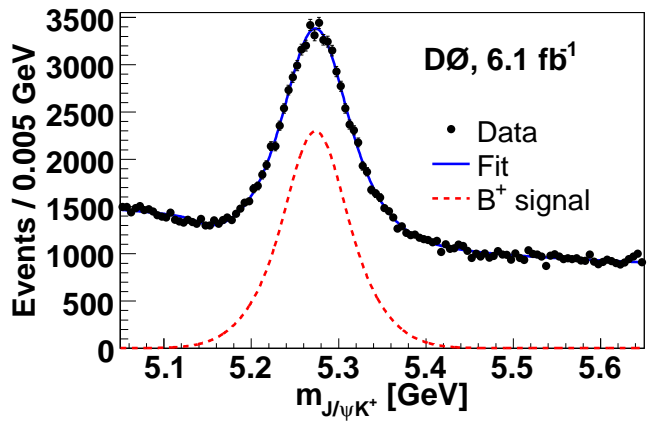}
  \caption{\label{fig:bumass} 
   The $J/\psi K^+$ invariant mass distribution of \BU\ candidates. The dashed line represents the $B^+$ signal distribution obtained from the fit (solid line).}
  \end{center}
\end{figure}

A sample of \BUJPSIK\ events is selected using all but the \BNN\ selection requirements, with an additional requirement of $p_T^{K}  \geq  1~{\rm GeV}$ for the kaon candidate.
By performing a binned likelihood fit with the $J/\psi K^+$ invariant mass distribution in data, we observe $N(B^+) = 14340 \pm 665$ events for \RUNIIA\ and $32463 \pm 875$ events for \RUNIIB, where the uncertainty is only statistical.
The statistical significance of the $B^+$ signal yield in \RUNIIB\ is higher than that in \RUNIIA\ although the lower yield per the integrated luminosity.
The $J/\psi K^+$ invariant mass distribution is shown in Fig.~\ref{fig:bumass}.
A systematic uncertainty of 2\% on the \BU\ yield is found by varying the fit parametrization.
The efficiency for the additional kaon track in \BUJPSIK\ decays
is calibrated using the ratio of \BDJPSIK\ to \BUJPSIK\ data with an uncertainty of 7.5\%.
The trigger efficiency depends on the muon transverse momentum $p_T^{\mu}$.
This is modeled by comparing the $p_T^{\mu}$ distribution in the selected data events with a control sample requiring a $p_T^{\mu}$ independent trigger and then applying the ratio to the simulated events as a $p_T^{\mu}$ dependent weight factor.
A possible dependence of this weight factor on the dimuon kinematics is evaluated
by choosing another sample at higher dimuon masses; this effect is found to be less than 1\%.
The $p_T^{B}$ spectra in the \BS\ and \BU\ simulations are corrected following comparisons of the \BUJPSIK\ in data and simulation.
A similar correction is obtained from \BSJPSIPHI\ decays, and the difference between the two is assigned as an uncertainty of 6.5\%.
The product of the factors multiplying $N(B_s^0)$ on the right-hand side of Eq.~\ref{eq:brcalc}
is called the single event sensitivity.
We find a single event sensitivity $(4.9 \pm 1.0)\times 10^{-9}$ for \RUNIIA\ and  $(1.84 \pm 0.36)\times 10^{-9}$ for \RUNIIB\ in the signal region.
Using the SM prediction of \BRBSMUMU~\cite{Buras:2009us}, there are $0.74 \pm 0.17$ events in \RUNIIA\ and $1.95 \pm 0.42$ events in \RUNIIB\ expected in the signal region.
Aside from the background uncertainty, the largest uncertainty of 15\% common to \RUNIIA\ and \RUNIIB\ comes from the fragmentation ratio, $f_u/f_s$.









We compute the final sensitivity using 2D histograms of $m_{\mu\mu}$~vs.~$\beta$ of the signal and the backgrounds by combining the sensitivity of each bin taking into account the correlated uncertainties.
In addition to the uncertainty on the signal normalization, we add uncertainties on the expected \BS\ mass and its resolution in the calculation.
Additional uncertainties on the dimuon background distributions are assigned
to allow for possible variation in the background $m_{\mu\mu}$ distribution as a function of \BNN.
The resulting median expected limits are \FULLEXPLIMITBRA\ for \RUNIIA, and \FULLEXPLIMITB\ for \RUNIIB\ at the 95\%~(90\%)~C.L.
and the combined median expected limit is \FULLEXPLIMITBR.
The limits are calculated from Eq.~\ref{eq:brcalc} using the semi-Frequentist confidence level approach (${\rm CL_{s}}$)~\cite{delphi_cls,Junk:1999kv,Fisher:2006zz}
with a Poisson log-likelihood ratio test statistic.
The limit incorporates Gaussian uncertainties on the signal efficiency and the background.
This expected limit is a factor of 2.4 better than the expected limit of $9.7 \times 10^{-8}$ at the 95\% C.L. of the previous D0 result~\cite{Abazov:2007iy},
where 10\% of this improvement results from changes in the analysis technique.

\begin{figure}
  \begin{center}
   \includegraphics[scale=0.43]{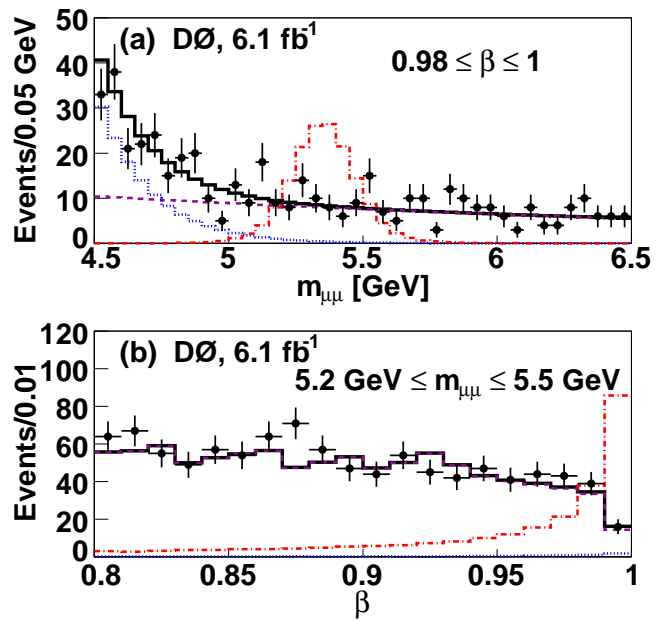}
  \caption{\label{fig:highsensitive} The distribution of $m_{\mu\mu}$ in the highest sensitivity $\beta$ region (a), and the distribution of $\beta$ in the highest sensitivity $m_{\mu\mu}$ region (b) for data (dots with uncertainties), expected background distribution (solid line), and the SM signal distribution multiplied by a factor of 100 (dotted-dashed line).
  The dimuon background contributions from the $B(D) \to \mu^+ \nu X, {\bar B}(\bar{D}) \to \mu^- \bar{\nu} X'$ decays (dashed line) and the $B \to \mu^+ \nu \bar{D}, \bar{D} \to \mu^- \bar{\nu} X$ decays (dotted line) are also shown.}
  \end{center}
\end{figure}

After finalizing the selection criteria and all systematic uncertainties, we study events in the signal region.
There are 256 events for \RUNIIA, and 823 events for \RUNIIB\ observed in the signal region where the expected number of background events is
$264 \pm 13$ events for \RUNIIA\ and $827 \pm 23$ events for \RUNIIB.
The observed distributions of dimuon events in the highest sensitivity region are shown in Fig.~\ref{fig:highsensitive}.
The observed number of events is consistent with the background expectations.
We extract 95\%~(90\%)~C.L. limits of \FULLOBSLIMITBRA\ for \RUNIIA\ and \FULLOBSLIMITB\ for \RUNIIB.
The resulting combined limit is \FULLOBSLIMITBR\ at the 95\%~(90\%)~C.L.
The probability for the expected background distributions to fluctuate to the observed data distributions is 31\%.

In conclusion, we have reported a search for the rare decay \BSMUMU\ using 6.1~\INVFB\ of \PPBAR\ collisions collected by the D0 experiment at Fermilab Tevatron Collider.
We observe no evidence for physics beyond the SM and set a limit of \FULLOBSLIMITBR\ at the 95\%~(90\%)~C.L.
This result is more stringent than the previous results~\cite{cdf:2007kv,Abazov:2007iy} and the best limit to date.
We thank the staffs at Fermilab and collaborating institutions,
and acknowledge support from the
DOE and NSF (USA);
CEA and CNRS/IN2P3 (France);
FASI, Rosatom and RFBR (Russia);
CNPq, FAPERJ, FAPESP and FUNDUNESP (Brazil);
DAE and DST (India);
Colciencias (Colombia);
CONACyT (Mexico);
KRF and KOSEF (Korea);
CONICET and UBACyT (Argentina);
FOM (The Netherlands);
STFC and the Royal Society (United Kingdom);
MSMT and GACR (Czech Republic);
CRC Program and NSERC (Canada);
BMBF and DFG (Germany);
SFI (Ireland);
The Swedish Research Council (Sweden);
and
CAS and CNSF (China).
%


\begin{thebibliography}{99}
\bibitem{charge_conjugation}
  Charge conjugate states are assumed implicitly throughout this paper.
\end{thebibliography}
\end{document}